\documentstyle[aps,preprint,floats,epsf,epsfig]{revtex}
\newcommand{\dgz}{\mbox{$\Delta g_{1}^{Z}\ $}}
\newcommand{\dkg}{\mbox{$\Delta \kappa_\gamma\ $}}
\newcommand{\lgm}{\mbox{$\lambda_\gamma\ $}}

\newcommand{\eratio}{\mbox{$\epsilon'/\epsilon \ $}}
\newcommand{\reeratio}{\mbox{$Re \ \epsilon'/\epsilon \ $}}

\newcommand{\be}{\begin{equation}}
\newcommand{\ee}{\end{equation}}
\newcommand{\bqa}{\begin{eqnarray}}
\newcommand{\eqa}{\end{eqnarray}}

\begin{document}

\author{
{\large\bf Francesco Terranova} \\
{\small Dipartimento di Fisica, Universit\`{a} di Milano-Bicocca \\
and INFN sez.~Milano, Italy }
}
%\date{}
% \phantom{xxx} \vspace{-1.0cm}
\title{
%{\normalsize\sf
%\rightline{}
%\rightline{}
%} 
{\normalsize\sf
\rightline{ }
\rightline{ }
\rightline{ }
}
\bigskip
{\LARGE\bf
The impact of LEP~2 data on possible anomalous enhancements
of \reeratio
}}

\maketitle
\thispagestyle{empty}

\phantom{xxx} \vspace{-6mm}

\begin{abstract}
It has been shown in the past that the real part of the 
\eratio ratio is particularly sensitive to anomalous
gauge couplings that modify the Standard Model Lagrangian. Due to the
loose bounds on these couplings coming from low energy processes and
to the poor sensitivity of hadron colliders to couplings such as \dgz,
it has been argued that anomalous couplings could still produce an
enhancement of $Re \ \epsilon' / \epsilon$ bringing this observable
closer to the experimental value obtained by KTeV, NA31 and NA48.
The impact of the new measurements done at LEP~2 in these years is discussed
and new severe constraints to this hypothesis are determined.
\end{abstract}

\newpage
\setcounter{page}{1}
\setcounter{footnote}{0}

%%%%%%%%%%%%%%%%%%%%%%%%%%%%%%%%%%%%%%%%%%%%%%%%%%%%%%%%%%%%%%%%%%%%%%%%%%%%%%%
% The main part of the paper
%%%%%%%%%%%%%%%%%%%%%%%%%%%%%%%%%%%%%%%%%%%%%%%%%%%%%%%%%%%%%%%%%%%%%%%%%%%%%%%

Since 1995, it has been noted \cite{he_95} that anomalous triple gauge
couplings (TGC) involving $WW\gamma$ and $WWZ$ vertices could modify
significantly the Standard Model (SM) prediction concerning \eratio. 
More precisely, strong penguin diagrams and isospin breaking due to quark
masses completely dominate the SM prediction of \reeratio 
for low values of $m_t$.
However, for $m_t$ of the order of 170~GeV, the effects of electroweak
penguin diagrams are sizeable \cite{lusignoli} and therefore \reeratio
becomes sensitive to anomalies in the bosonic sector of the SM.

The possibility that anomalous TGC could modify and, particularly, enhance
\reeratio has been considered with interest, especially because several
experimental measurements still point towards a rather high value of \reeratio
with respect to the SM expectation. In fact,  present theoretical 
errors on \reeratio, coming mainly from the uncertainties in the 
models for hadronic matrix elements, prevent us from drawing conclusions
about the effectiveness of SM in the CP violating sector. On the other
hand, it is interesting to quantify to what extent the hypothesis of anomalous 
enhancement of \reeratio is corroborated by the intense experimental 
investigation on the bosonic sector of the SM carried out at LEP~2 since 1996.

It is customary to express general deviations from the SM in the bosonic
sector in the framework of effective theories \cite{effective}.
In this case, the most general Lagrangian invariant under $U_{em}(1)$ 
that contributes to $WWV$ vertices ($V=Z,\gamma$)
\cite{gaemers,HHPZ} is:
\be
\begin{array}{ll}
 i{\cal L}_{eff}^{WWV} = &g_{WWV} \cdot [  g_1^V V^\mu
 (W^-_{\mu\nu}W^{+\nu} - W^+_{\mu\nu}W^{-\nu}) +
 \kappa_VW^+_\mu W^-_\nu V^{\mu\nu} +   \label{lagratgc} \\
 ~&+ \frac{\lambda_V}{M_W^2}V^{\mu\nu}W^{+\rho}_\nu W^-_{\rho\nu} +
 ig_5^V\varepsilon_{\mu\nu\rho\sigma}((\partial^\rho W^{-\mu})W^{+\nu} -
                                 W^{-\mu}(\partial^\rho W^{+\nu})) V^\sigma +
\\ ~&
 + ig_4^VW^-_\mu W^+_\nu(\partial^\mu V^\nu + \partial^\nu V^\mu) -
 \frac{\tilde{\kappa}_V}{2}W^-_\mu
 W^+_\nu\varepsilon^{\mu\nu\rho\sigma}V_{\rho\sigma} -
 \frac{\tilde{\lambda}_V}{2M_W^2} W^-_{\rho\mu}W^{+\mu}_\nu
 \varepsilon^{\nu\rho\alpha\beta}V_{\alpha\beta} ] \ \ , 
\end{array}
\label{genlag}
\ee
\noindent where $W^{\pm \mu}$ are the $W$ boson fields and $V=\gamma,Z$.
Defining
$g_{WW\gamma}=e$ and $g_{WWZ}=e \ cot \theta_W $, 
this Lagrangian allows anomalous
values for the C- and P-conserving couplings 
$\kappa_\gamma$, $\kappa_Z$, $g_1^Z$, $g_1^\gamma$
(equal to 1 in SM).
Moreover, new contributions coming from operators absent in the standard theory
are present. These are $\lambda_V$, which also
conserves both $C$ and $P$-parity; $g_5^V$
($C$ and $P$ violating but $CP$ conserving) and the $CP$ violating terms
$g_4^V$, ${\tilde \kappa_V}$ and ${\tilde \lambda_V}$.
In terms of the $W$ magnetic dipole and electric quadrupole
they contribute as
\bqa
\mu_W = (g_1^\gamma+\kappa_\gamma+\lambda_\gamma) \frac{e}{2m_W} s \ ; &
Q_W = - \frac{e ( \kappa_\gamma-\lambda_\gamma) }{m_W^2} \ \ , \\ \nonumber 
\eqa
where $eg_\gamma^1$ is the $W$ charge and $s$ its spin.
To simplify the notations, it is convenient to introduce the
deviation from the SM couplings
\begin{center}
$ \Delta g_1^V \equiv g_1^V-1 \ \ , \ \  
\Delta \kappa_V \equiv \kappa_V - 1 \ \ .$ 
\end{center}

\noindent
CP-violating terms are constrained by the neutron and electron electric
dipole moments \cite{neutron}. CP-conserving anomalous couplings
affect low-energy rare decays \cite{lowE}, $W$ production processes at 
high scales (LEP~2 and Tevatron) and electroweak corrections to the $W$,$Z$,
$\gamma$  propagator (``oblique corrections''). Oblique corrections have 
been extensively tested at the scale of $m_Z$ at LEP~1 and SLC. 
In order to evade the tight constraints already obtained from LEP at the 
$Z^0$-pole, present
LEP~2 measurements are expressed as limits to couplings contributing to the
effective Lagrangian (\ref{genlag}) after imposing $SU(2) \otimes U(1)$ 
gauge  invariance and retaining only the lowest dimension operators 
\cite{derujula,HISZ,yellow1}. This approach implies relations amongst
the various TGC \cite{yellow1} and reduce the number of independent 
couplings to three: \dgz, \lgm and \dkg. 

The CP-violating $\Delta S=1$ interaction responsible for 
$K \rightarrow \pi \pi$ is affected, beyond strong penguins, by electroweak
penguins and (possibly) TGC contributions changing both the $I=0$ and $I=2$
amplitude.
At the scale of $\mu=m_W$, the SM effective Hamiltonian ($H_{eff}$) 
for $\Delta S$=1, modified in order to cope with anomalous TGC, 
has been computed in \cite{he}.
It has been shown that the CP-violating couplings and $\Delta \kappa_Z$ 
do not contribute to leading order, being suppressed by factors of
$O((m_{d,s}^2, m_K^2)/m_W^2)$. Running $H_{eff}$ at the scale of $\mu=1$~GeV  
implies the knowledge of the boundary conditions of the Wilson coefficients
in SM \cite{13_he} and in the occurrence of anomalous couplings \cite{he}.
In the calculation of \cite{he} a cut-off $\Lambda$ of 1~TeV has been used
for terms proportional to \dgz and \dkg. 
The results obtained in \cite{he} depend, in general, on the anomalous 
couplings \dgz,\dkg, \lgm and $g_5^Z$, on the imaginary part of 
$V_{td}V^*_{ts}$
and on the assumptions about the hadronic matrix elements. 

%Relying on 
%\cite{18_he} for predictions concerning the hadronic matrix elements,
%the following expression has been derived for \reeratio in
%terms of the remaining variables:
The dependence of \reeratio to anomalous TGC can be expressed in the
following way:

\be
Re \left( \frac{\epsilon'}{\epsilon} \right) \ 
\simeq \ {\cal H} \ Im (V_{td}V^*_{ts}) \ (1+0.96\dkg+0.16\lgm
-4.08 \dgz +0.44 g_5^Z) \ .
\label{redep}
\ee

\noindent where the overall normalisation factor ${\cal H}$ depends
mainly on the knowledge of the hadronic matrix elements and, according
to the calculation of \cite{18_he}, ${\cal H} \simeq 8.66$.
\noindent Due to the clear dominance of the term involving \dgz, 
Tevatron results do not modify 
significantly the low-energy constraints and are not considered here.
In the present analysis, we take the current allowed range of 
$Im (V_{td}V^*_{ts})$ from \cite{bertolini_2000} and we assume

\be
Im (V_{td}V^*_{ts}) = (1.14\pm 0.11) \cdot 10^{-4}
\ee

\begin{figure}[htb]
\centerline{\epsfig
{file=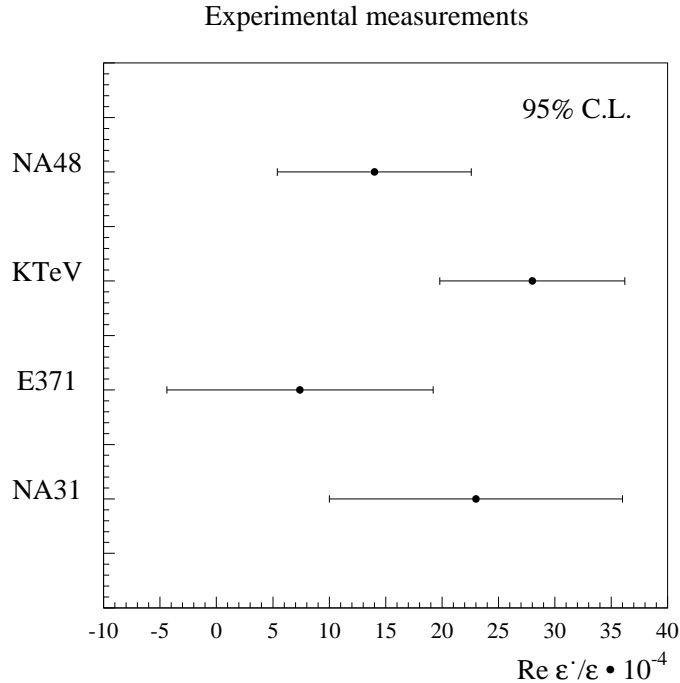,height=10cm} }
\caption{
Present experimental measurements of \reeratio (95\% C.L.). }
\label{exp} 
\end{figure}

It is well known that the status of the experimental  measurement 
of \reeratio is not completely satisfactory. First results 
were obtained by NA31 \cite{NA31} and were not confirmed by
E731 \cite{E731}. Preliminary results from the KTeV \cite{KTeV} collaboration
and from NA48 \cite{NA48} 
strongly support a non-zero value of \reeratio (fig.\ref{exp}).
On the theoretical side, current calculations of 
\reeratio in SM point, in general, towards lower values than those suggested 
by current experiments, even if the statistical significance is of the order
of 2-2.5$\sigma$ (for a review see \cite{he_1,bertolini_2000}) and
make the hypothesis of an anomalous enhancement of \reeratio due to TGC
rather attractive. 
It is interesting to note, however, that some very recent 
updates of the theoretical calculations of the hadronic
matrix elements \cite{bertolini_2000,buras} 
could imply a higher value of \reeratio which, if
confirmed, would ease the agreement of the SM predictions with
the current experimental measurements without invoking new physics.
On the other hand, for what concerns TGC the hadronic matrix elements
mainly affects the overall normalisation factor ${\cal H}$ being
the relative contributions
of the couplings (last factor of eq.(\ref{redep}) ) practically independent
to them.
Therefore, the allowed ranges derived in the following 
assuming ${\cal H} \simeq 8.66$ can be updated in a straightforward 
manner by proper rescaling of ${\cal H}$ in eq.(\ref{redep}).

\begin{figure}[htb]
\centerline{\epsfig
{file=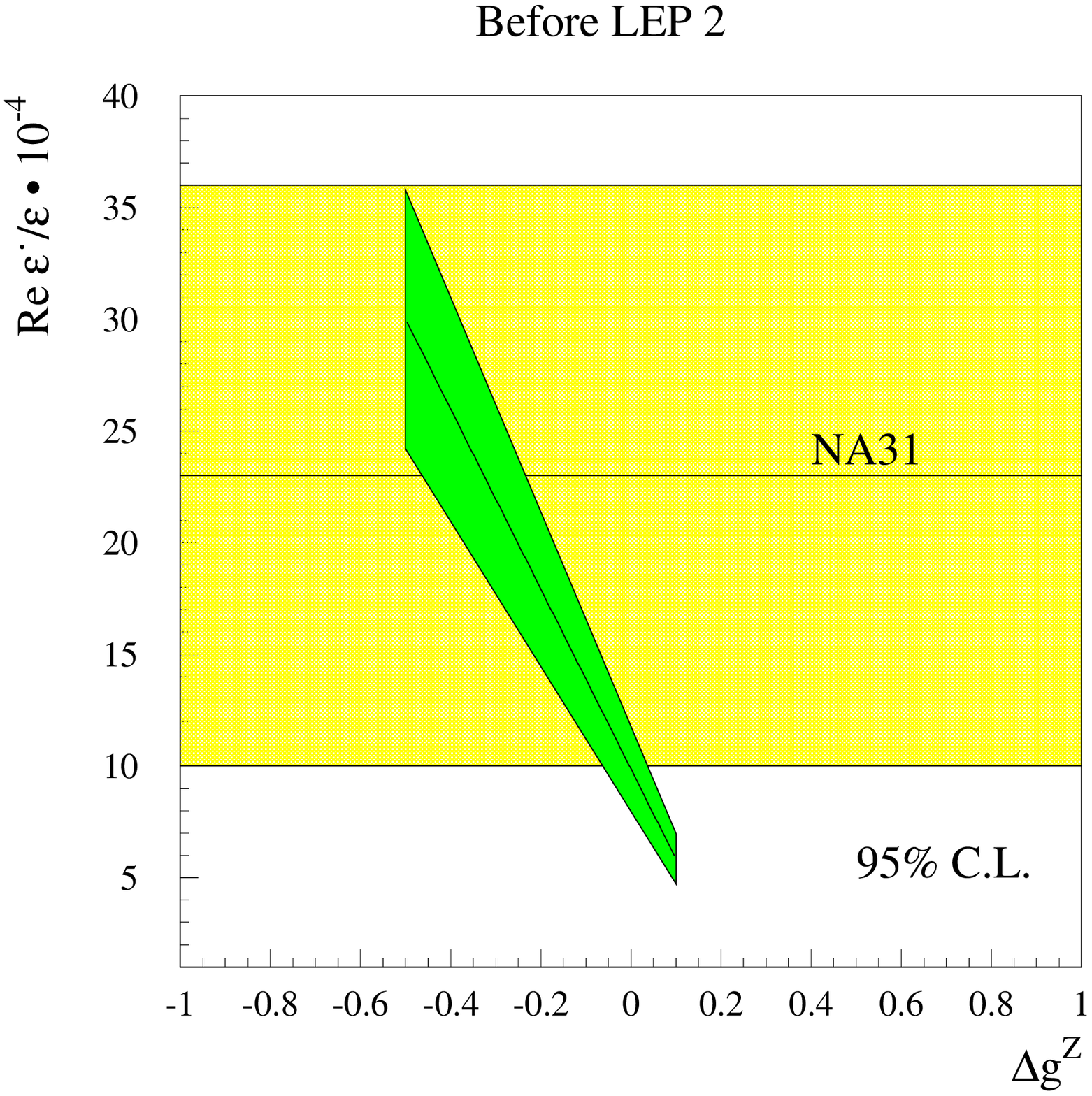,height=10cm} }
\caption{
\reeratio as a function of \dgz for the range of this coupling allowed by
low energy data (fixing all the couplings but \dgz to their SM values).
The vertical width of the dark band represents the variation 
of  \reeratio corresponding to a change of 
$\pm 2 \sigma$ of $Im (V_{td}V^*_{ts})$. 
The horizontal extent of the dark band represents
the allowed range of \dgz.
The light band indicates the allowed 
range from NA31. All the limits are computed at 95\% C.L. }
\label{beflep2} 
\end{figure}

Figure \ref{beflep2} shows the possible enhancement of \reeratio as a function
of \dgz, without including current LEP~2 data and assuming all TGC but \dgz
at their SM values. 
The allowed range of \dgz has been computed including just the low 
energy constraints coming from rare $B$ and $K$ decay, as in \cite{he}. 
The vertical width of the dark band represents the variation of 
\reeratio corresponding to a change of $\pm 2 \sigma$ of $Im (V_{td}V^*_{ts})$.
The light band indicates the allowed range from NA31.
All the limits are computed at 95\% C.L. The  plot represents approximately 
the experimental situation at the beginning of the high energy run of LEP.

\begin{figure}[htb]
\centerline{\epsfig
{file=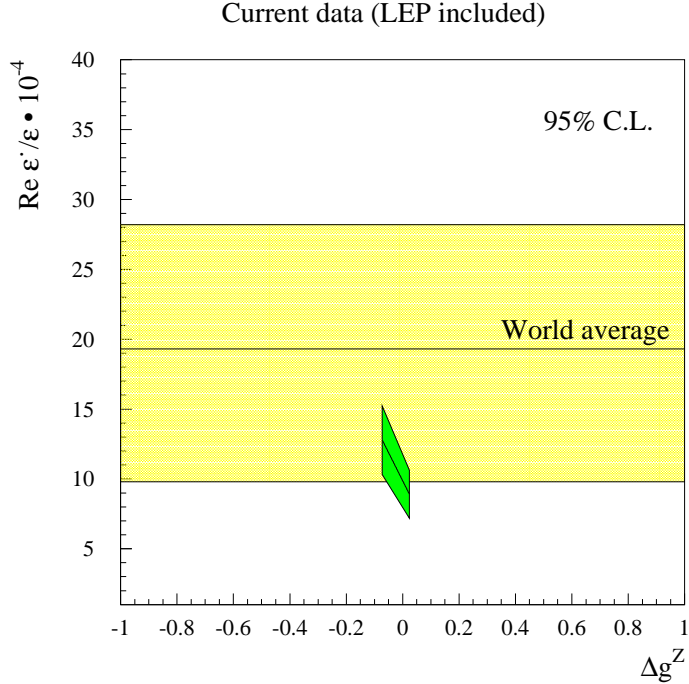,height=10cm} }
\caption{
\reeratio as a function of \dgz for the range of this coupling allowed by
low energy and LEP data (fixing all the couplings but \dgz to their SM values).
The vertical width of the dark band represents the variation 
of  \reeratio corresponding to a change of 
$\pm 2 \sigma$ of $Im (V_{td}V^*_{ts})$. 
The horizontal extent of the dark band represents
the allowed range of \dgz.
The light band indicates the current
world average of \reeratio (see text for details). 
All the limits are computed at 95\% C.L. }
\label{aftlep2} 
\end{figure}

The current situation is depicted in figure \ref{aftlep2}. 
Here the world average
has been computed inflating the error until $\chi^2/ndf=1$, 
in order to deal with
the discrepancy between the old E731 measurement and the analyses
of KTeV, NA31 and NA48. Simple rejection of E731 results in a world average
of $( \ 21.6 \pm 2.7 \ ) \cdot 10^{-4} $.
The allowed range for \dgz combines the 95\% C.L. limits coming from
ALEPH, DELPHI, OPAL and L3 \cite{lepdata} summing up the accumulated
statistics from 1996 to 1999 taken 
at centre-of-mass energies ranging from 161~GeV to 202~GeV. 
Here, statistical  and systematic uncertainties are included.

The remarkable improvement in the experimental tests of anomalous values
of \dgz strongly constraints the hypothesis of anomalous enhancement, bringing
the allowed range of \reeratio to

\be
8.0 \ 10^{-4} < \reeratio < 13.7 \ 10^{-4} \ \ \ (95\% \ C.L.) \ ,
% Tampere
%7.1 \ 10^{-4} < \reeratio < 13.4 \ 10^{-4} \ \ \ (95\% \ C.L.) \ ,
%6.6 \ 10^{-4} < \reeratio < 14.6 \ 10^{-4} \ \ \ 95\% \ C.L.
\ee

\noindent to be compared with the pre-LEP~2 measurement of

\be
5.4 \ 10^{-4} < \reeratio < 30.5 \ 10^{-4} \ \ \ (95\% \ C.L.) \ .
%4.8 \ 10^{-4} < \reeratio < 35.8 \ 10^{-4} \ \ \ 95\% \ C.L.
\ee

\noindent These limits have been extracted assuming that only \dgz is different
from zero. In fact, single $W$ production at LEP~2 strongly constraints 
contributions to the effective Lagrangian coming 
from operators proportional to \dkg. The accumulated LEP~2 statistics
allows a simultaneous fit of \dgz and \dkg (again, all other couplings have
been fixed to their SM value).  The corresponding allowed range for
\reeratio is

\be
7.9 \ 10^{-4} < \reeratio < 15.6 \ 10^{-4} \ \ \ 95\% \ C.L.
%3.5 \ 10^{-4} < \reeratio < 17.1 \ 10^{-4} \ \ \ 95\% \ C.L.
\ee

Contributions from other couplings have
very limited impact on \reeratio by virtue of eq.(\ref{redep}).
It has to be noted that results for the TGC parameters have been quoted
without use of a form factor $\Lambda$, describing the 
scale at which new physics
should become manifest. Inclusion of such a parameter with $\Lambda =1$~TeV,
for sake of consistency with \cite{he}, and form factors of the type used, 
for example, in  \cite{aihara} 
would increase the limits obtained for \reeratio by no more than 4\%.

%\begin{figure}[htb]
%\centerline{\epsfig
%{file=lepcomb1.eps,width=10cm} }
%\caption{
%Current limits on (\dgz,\dkg) from LEP~2 data (from \cite{lepdata}).}
%\label{limits2d} 
%\end{figure}

In conclusion, we have shown that present LEP~2 data highly constrain
the hypothesis of a non-standard enhancement of \reeratio coming from 
anomalies in the bosonic sector of the SM. In particular, the sensitivity
of the process $e^+e^- \rightarrow W^+W^-$ to \dgz 
allows a reduction of the possible contribution to \reeratio
from this coupling by a factor of about~4.

%allowed a reduction of the
%possible range of this coupling, dominating the anomalous 
%contribution to \reeratio, of about a factor 6.

{\bf{Acknowledgements} } \\
I'm greatly indebted to X.-G.~He, C.~Matteuzzi and R.~Sekulin for useful 
discussions  on this subject. 
Interesting suggestions of R.~Felici and V.~Verzi are
gratefully acknowledged.

%%%%%%%%%%%%%%%%%%%%%%%%%%%%%%%%%%%%%%%%%%%%%%%%%%%%%%%%%%%%%%%%%%%%%%%%%%
%BIBLIOGRAPHY
%%%%%%%%%%%%%%%%%%%%%%%%%%%%%%%%%%%%%%%%%%%%%%%%%%%%%%%%%%%%%%%%%%%%%%%%%%

\vfill\eject

\end{document}